\documentclass[12pt,preprint]{aastex}
\usepackage{epsfig}
\usepackage{graphicx}
\usepackage[dvips]{color}
\usepackage{ulem}

\newcommand{\es}{1ES\,1959+650}
\newcommand{\swift}{\textit{Swift} }
\newcommand{\fermi}{\textit{Fermi} }

\slugcomment{ApJ accepted October 1 2014}

\shorttitle{1ES 1959+650 Flare}
\shortauthors{Furniss et al.}

\begin{document}

\title{Investigating Broadband Variability of the TeV Blazar \es}

\author{
E.~Aliu\altaffilmark{1},
S.~Archambault\altaffilmark{2},
T.~Arlen\altaffilmark{3},
T.~Aune\altaffilmark{3},
A.~Barnacka\altaffilmark{4},
M.~Beilicke\altaffilmark{5},
W.~Benbow\altaffilmark{6},
K.~Berger\altaffilmark{7},
R.~Bird\altaffilmark{8},
A.~Bouvier\altaffilmark{9},
J.~H.~Buckley\altaffilmark{5},
V.~Bugaev\altaffilmark{5},
M.~Cerruti\altaffilmark{6},
X.~Chen\altaffilmark{10,11},
L.~Ciupik\altaffilmark{12},
E.~Collins-Hughes\altaffilmark{8},
M.~P.~Connolly\altaffilmark{13},
W.~Cui\altaffilmark{14},
J.~Dumm\altaffilmark{15},
J.~D.~Eisch\altaffilmark{16},
A.~Falcone\altaffilmark{17},
S.~Federici\altaffilmark{11,10},
Q.~Feng\altaffilmark{14},
J.~P.~Finley\altaffilmark{14},
H.~Fleischhack\altaffilmark{11},
P.~Fortin\altaffilmark{6},
L.~Fortson\altaffilmark{15},
A.~Furniss\altaffilmark{9},
N.~Galante\altaffilmark{6},
G.~H.~Gillanders\altaffilmark{13},
S.~Griffin\altaffilmark{2},
S.~T.~Griffiths\altaffilmark{18},
J.~Grube\altaffilmark{12},
G.~Gyuk\altaffilmark{12},
N.~H{\aa}kansson\altaffilmark{10},
D.~Hanna\altaffilmark{2},
J.~Holder\altaffilmark{7},
G.~Hughes\altaffilmark{11},
T.~B.~Humensky\altaffilmark{19},
C.~A.~Johnson\altaffilmark{9},
P.~Kaaret\altaffilmark{18},
P.~Kar\altaffilmark{20},
M.~Kertzman\altaffilmark{21},
Y.~Khassen\altaffilmark{8},
D.~Kieda\altaffilmark{20},
H.~Krawczynski\altaffilmark{5},
F.~Krennrich\altaffilmark{16},
M.~J.~Lang\altaffilmark{13},
A.~S~Madhavan\altaffilmark{16},
P.~Majumdar\altaffilmark{3,22},
S.~McArthur\altaffilmark{23},
A.~McCann\altaffilmark{24},
K.~Meagher\altaffilmark{25},
J.~Millis\altaffilmark{26,26},
P.~Moriarty\altaffilmark{27,13},
R.~Mukherjee\altaffilmark{1},
T.~Nelson\altaffilmark{15},
D.~Nieto\altaffilmark{19},
A.~O'Faol\'{a}in de Bhr\'{o}ithe\altaffilmark{11},
R.~A.~Ong\altaffilmark{3},
A.~N.~Otte\altaffilmark{25},
N.~Park\altaffilmark{23},
J.~S.~Perkins\altaffilmark{28},
M.~Pohl\altaffilmark{10,11},
A.~Popkow\altaffilmark{3},
H.~Prokoph\altaffilmark{11},
J.~Quinn\altaffilmark{8},
K.~Ragan\altaffilmark{2},
J.~Rajotte\altaffilmark{2},
L.~C.~Reyes\altaffilmark{29},
P.~T.~Reynolds\altaffilmark{30},
G.~T.~Richards\altaffilmark{25},
E.~Roache\altaffilmark{6},
A.~Sadun\altaffilmark{32},
M.~Santander\altaffilmark{1},
G.~H.~Sembroski\altaffilmark{14},
K.~Shahinyan\altaffilmark{15},
F.~Sheidaei\altaffilmark{20},
A.~W.~Smith\altaffilmark{20},
D.~Staszak\altaffilmark{2},
I.~Telezhinsky\altaffilmark{10,11},
M.~Theiling\altaffilmark{14},
J.~Tyler\altaffilmark{2},
A.~Varlotta\altaffilmark{14},
V.~V.~Vassiliev\altaffilmark{3},
S.~Vincent\altaffilmark{11},
S.~P.~Wakely\altaffilmark{23},
T.~C.~Weekes\altaffilmark{6},
A.~Weinstein\altaffilmark{16},
R.~Welsing\altaffilmark{11},
A.~Wilhelm\altaffilmark{10,11},
D.~A.~Williams\altaffilmark{9},
B.~Zitzer\altaffilmark{31}
 (The VERITAS Collaboration)
and
M.~B\"ottcher\altaffilmark{*,33}, M.~Fumagalli\altaffilmark{34,35,36}
}

\altaffiltext{*}{Corresponding authors: amy.furniss@gmail.com, Markus.Bottcher@nwu.ac.za}
\altaffiltext{1}{Department of Physics and Astronomy, Barnard College, Columbia University, NY 10027, USA}
\altaffiltext{2}{Physics Department, McGill University, Montreal, QC H3A 2T8, Canada}
\altaffiltext{3}{Department of Physics and Astronomy, University of California, Los Angeles, CA 90095, USA}
\altaffiltext{4}{Harvard-Smithsonian Center for Astrophysics, 60 Garden Street, Cambridge, MA 02138, USA}
\altaffiltext{5}{Department of Physics, Washington University, St. Louis, MO 63130, USA}
\altaffiltext{6}{Fred Lawrence Whipple Observatory, Harvard-Smithsonian Center for Astrophysics, Amado, AZ 85645, USA}
\altaffiltext{7}{Department of Physics and Astronomy and the Bartol Research Institute, University of Delaware, Newark, DE 19716, USA}
\altaffiltext{8}{School of Physics, University College Dublin, Belfield, Dublin 4, Ireland}
\altaffiltext{9}{Santa Cruz Institute for Particle Physics and Department of Physics, University of California, Santa Cruz, CA 95064, USA}
\altaffiltext{10}{Institute of Physics and Astronomy, University of Potsdam, 14476 Potsdam-Golm, Germany}
\altaffiltext{11}{DESY, Platanenallee 6, 15738 Zeuthen, Germany}
\altaffiltext{12}{Astronomy Department, Adler Planetarium and Astronomy Museum, Chicago, IL 60605, USA}
\altaffiltext{13}{School of Physics, National University of Ireland Galway, University Road, Galway, Ireland}
\altaffiltext{14}{Department of Physics and Astronomy, Purdue University, West Lafayette, IN 47907, USA}
\altaffiltext{15}{School of Physics and Astronomy, University of Minnesota, Minneapolis, MN 55455, USA}
\altaffiltext{16}{Department of Physics and Astronomy, Iowa State University, Ames, IA 50011, USA}
\altaffiltext{17}{Department of Astronomy and Astrophysics, 525 Davey Lab, Pennsylvania State University, University Park, PA 16802, USA}
\altaffiltext{18}{Department of Physics and Astronomy, University of Iowa, Van Allen Hall, Iowa City, IA 52242, USA}
\altaffiltext{19}{Physics Department, Columbia University, New York, NY 10027, USA}
\altaffiltext{20}{Department of Physics and Astronomy, University of Utah, Salt Lake City, UT 84112, USA}
\altaffiltext{21}{Department of Physics and Astronomy, DePauw University, Greencastle, IN 46135-0037, USA}
\altaffiltext{22}{Saha Institute of Nuclear Physics, Kolkata 700064, India}
\altaffiltext{23}{Enrico Fermi Institute, University of Chicago, Chicago, IL 60637, USA}
\altaffiltext{24}{Kavli Institute for Cosmological Physics, University of Chicago, Chicago, IL 60637, USA}
\altaffiltext{25}{School of Physics and Center for Relativistic Astrophysics, Georgia Institute of Technology, 837 State Street NW, Atlanta, GA 30332-0430}
\altaffiltext{26}{Department of Physics, Anderson University, 1100 East 5th Street, Anderson, IN 46012}
\altaffiltext{27}{Department of Life and Physical Sciences, Galway-Mayo Institute of Technology, Dublin Road, Galway, Ireland}
\altaffiltext{28}{N.A.S.A./Goddard Space-Flight Center, Code 661, Greenbelt, MD 20771, USA}
\altaffiltext{29}{Physics Department, California Polytechnic State University, San Luis Obispo, CA 94307, USA}
\altaffiltext{30}{Department of Applied Physics and Instrumentation, Cork Institute of Technology, Bishopstown, Cork, Ireland}
\altaffiltext{31}{Argonne National Laboratory, 9700 S. Cass Avenue, Argonne, IL 60439, USA}
\altaffiltext{32}{Department of Physics, University of Colorado Denver, CO, USA}
\altaffiltext{33}{Centre for Space Research, Private Bag X6001, North-West University, Potchefstroom Campus, Potchefstroom, 2520, South Africa}
\altaffiltext{34}{Carnegie Observatories, 813 Santa Barbara Street, Pasadena, CA 91101, USA.}
\altaffiltext{35}{Department of Astrophysics, Princeton University, Princeton, NJ 08544-1001, USA.}
\altaffiltext{36}{Hubble Fellow}

\email{amy.furniss@gmail.com}
\email{markus.bottcher@nwu.ac.za}

\begin{abstract}
We summarize broadband observations of the TeV-emitting blazar \es, including optical R-band observations by the robotic telescopes Super-LOTIS and iTelescope, UV observations by \swift UVOT, X-ray observations by the \swift X-ray Telescope (XRT), high-energy gamma-ray observations with the \textit{Fermi} Large Area Telescope (LAT) and very-high-energy (VHE) gamma-ray observations by VERITAS above 315 GeV, all taken between 17 April 2012 and 1 June 2012 (MJD 56034 and 56079).  The contemporaneous variability of the broadband spectral energy distribution is explored in the context of a simple synchrotron self Compton (SSC) model.  In the SSC emission scenario, we find that the parameters required to represent the high state are significantly different than those in the low state.  Motivated by possible evidence of gas in the vicinity of the blazar, we also investigate a reflected-emission model to describe the observed variability pattern.  This model assumes that the non-thermal emission from the jet is reflected by a nearby cloud of gas, allowing the reflected emission to re-enter the blob and produce an elevated gamma-ray state with no simultaneous elevated synchrotron flux.  The model applied here, although not required to explain the observed variability pattern, represents one possible scenario which can describe the observations.  As applied to an elevated VHE state of 66\% of the Crab Nebula flux, observed on a single night during the observation period, the reflected-emission scenario does not support a purely leptonic non-thermal emission mechanism.   The reflected emission model does, however, predict a reflected photon field with sufficient energy to enable elevated gamma-ray emission via pion production with protons of energies between 10 and 100 TeV.
\end{abstract}

  
\section{Introduction}
\es, a blazar at $z$=0.047 \citep{scatcher}, was among the first-detected extragalactic very-high-energy (VHE; $E\ge$100 GeV) emitters \citep{holder1959}.  This blazar has recently been observed by VERITAS at approximately $23\%$ Crab Nebula flux\footnote{The Crab Nebula is a bright VHE source, roughly characterized with a power-law index of $\Gamma$=2.5 in the VHE band, where $dN/dE\propto(E)^{-\Gamma}$.  
This source has an integral flux of approximately 1.2$\times10^{-10}$ ph cm$^{-2}$s$^{-1}$ above 315 GeV, according to the fit to the VHE data in \cite{albertCrab}.} above 1 TeV \citep{anna}.  A blazar is a type of active galactic nucleus (AGN) having a relativistic jet that is oriented close to the line of sight of the observer.  The non-thermal radiation from blazars, thought to originate from within the jet, produces a double-humped spectral energy distribution (SED). 

The lower-energy component of the SED, referred to as the synchrotron component, results from the synchrotron radiation of relativistic leptons in the presence of a tangled magnetic field.  The higher-energy component, hereafter referenced as the gamma-ray component, can be attributed to inverse-Compton up-scattering by the relativistic particles within the jet of either the synchrotron photons themselves (synchrotron self-Compton emission: SSC) or a photon field external to the jet (external-Compton emission: EC).  The photon fields may arise from emission by the accretion disk, a broad line region or a dusty torus, as described in  \cite{dermer}, \cite{maraschi}, \cite{marscher} and \cite{sikora}.  Hadronic processes initiated by relativistic protons (such as pion production and the resulting cascade emission) can similarly produce a gamma-ray component \citep{aharonian02,bednarek,dar,mannheim,mucke,pohl}. 

The non-thermal emission resulting from these different processes can produce nearly indistinguishable time-averaged SEDs, as discussed in \cite{tagliaferri}, making emission-model discrimination based on non-simultaneous data uncertain.  An effective means to investigate blazar emission mechanisms is through the observation and subsequent modeling of broadband spectral variability \citep{coppi,bot07,krawc2002}.  There is some evidence that for many objects the low-energy and high-energy peaks vary in concert (e.g. Mrk 421: \cite{fossati}, Mrk 501: \cite{krawc2000}).  Correlated variability between the low and high-energy SED components can be well described by a simple SSC model, whereas less-common, uncorrelated variability patterns, similar to the ``orphan" flaring event observed from \es\,  \citep{krawczynski}, require more complex emission scenarios.  This type of non-correlated variability has been described using multiple-zone SSC emission, EC emission, a model including a magnetic field aligned along the jet axis, hadronic emission and reflected emission scenarios \citep{krawczynski,graff,kusunose,boettcher05}.

The central engines of AGNs have multiple components, possibly including various gas and dust tori and clouds. These clouds can scatter some of the continuum emission from the accretion disk and from the jet.  As this process may be common to AGN, although at different scales, we include the evaluation of reflected-emission in the theoretical discussion of our results.  A reflected-emission scenario requires the non-thermal emission region to be in close proximity to a dilute gas.  In this paradigm, the gas reflects synchrotron emission via Thompson scattering back to the non-thermal emission region, providing an external photon field to be up-scattered.  The broadband variability pattern resulting from such an emission geometry would be displayed as an elevated state of the gamma-ray component alone. 

Evidence from millimeter \citep{fumagalli} and X-ray observations \citep{furniss} of \es\, support the existence of intervening gas within the blazar and, therefore, the application of a reflected-emission scenario to broadband variability of the source.  More specifically, 1ES\,1959+650 shows evidence for additional gas in the vicinity of the host galaxy from X-ray absorption in excess to that expected by the Galactic column, as well as a positive detection of molecular CO within the blazar.  

In this work we summarize broadband variability of \es\, detected during multiwavelength observations between 17 April 2012 and 1 June 2012 (MJD 56034 and 56079).  These observations include 0.7 ks of strictly simultaneous VERITAS and \swift observations on MJD 56067, occurring at the beginning of a VHE flare lasting approximately two hours.  A simple SSC emission scenario is applied to the data to investigate which parameter changes are required to produce the observed variability.  Motivated by the recent evidence for intervening gas within \es, we also investigate a possible explanation of the broadband variability through the application of a reflected-emission scenario.  We summarize the multiwavelength observations and analysis in Sections 2.  The models are applied to the data in Section 3, and the implications of each model application are discussed in Section 4.  Throughout this work, the term ``flare`` is used to denote an elevated state of at least five times the average flux measured over the period of observations, with at least a $5\sigma$ deviation from the average (calculated without inclusion of any ``flaring" state.)

\section{Broadband Observations}

\subsection{VERITAS}
VERITAS is an array of four imaging atmospheric Cherenkov telescopes in southern Arizona, each with a 3.5$^{\circ}$ field of view.  The array is sensitive to photons with energies from $\sim$100 GeV to more than 10 TeV and can detect a 1\% Crab-Nebula-flux source at 5 standard deviations ($\sigma$) in less than 28 hours.  The telescope array uses 12-meter reflectors to focus dim, blue/UV Cherenkov light from gamma-ray and cosmic-ray interactions in the atmosphere onto cameras composed of 499 photomultiplier tubes (PMTs).  More details on the VERITAS instrument can be found in \cite{holder} and \cite{weekes}.

A historically high optical state of \es\, (R-band of 13.8 magnitude, as measured by the Super-LOTIS robotic telescope; \citealt{superlotis}), prompted near-nightly VERITAS exposures over two dark runs\footnote{A dark run is the period between two consecutive full moons.} between 17 April 2012 and 1 June 2012 (MJD 56034-56079).  Observations were carried out in \textit{wobble} mode, with exposures taken at 0.5$^{\circ}$ offset in each of the four cardinal directions from \es, in order to facilitate simultaneous background measurements \citep{fomin, berge}. The total exposure over the two dark runs resulted in 8.7 hours of quality-selected live time which was collected at an average zenith angle of 37$^{\circ}$.

Elliptical moments of the recorded images are calculated and used to discriminate background cosmic-ray events from gamma-ray events.  The data are first cleaned with ``quality cuts", discarding any telescope images involving fewer than five PMTs or images with centroids at greater than 1.43$^{\circ}$ from the camera center.  Additionally, each image is required to have a total ``size" (a measure of total Cherenkov light collected by the camera) of more than $\sim$80 photoelectrons. 

Single-telescope image widths and lengths are combined into mean-scaled-width (MSW) and mean-scaled-length (MSL) parameters for each array event, as described in \cite{cogan} and \cite{daniel}.   Only array events with 0.05 $<$ MSW $<$1.25, 0.05 $<$ MSL $<$1.1, a reconstructed height of shower maximum greater than 7 km above the array, and having a reconstruction direction within 0.1$^{\circ}$ of \es\, are kept as candidate signal (ON) events.  The background (OFF) events are those which pass all aforementioned cuts and fall within 0.1$^{\circ}$-radius circular regions at the same radial distance as the source from the center of the camera.  The source significance is calculated from the number of events falling in these ON and OFF regions according to Equation 17 of \cite{lima}.  The analysis of the VHE source signal is confirmed with two independent analysis packages.

The total VERITAS exposure between 17 April 2012 and 1 June 2012 (MJD 56034 to 56079) resulted in 517 ON events and 1175 OFF events (corresponding to 410 excess events with an averaged background normalization $\alpha$ parameter of 0.0909) and an overall detection significance of 31$\sigma$.  The VERITAS spectral data are derived with systematically coarser binning with increasing energy and are fit with a differential power law of the form $dN/dE=N_0 \times (E/E_0)^{-\Gamma}$, where $E_0$ is fixed at  1 TeV and $N_0$ is the normalization parameter.  Variability was detected during these observations, as can be seen by the source spectra in Figure 1 and in the top panel of the broadband light curve presented in Figure 2. The upper limits represent 95\% confidence upper limits, calculated according to \cite{rolke}.  A $\chi$-squared test shows less than 6.4$\times10^{-12}$ probability of a steady VHE flux.   Table 1 contains a summary of the VHE analysis and spectral states of \es\, during the two dark runs (MJD 56034-56040 and 56064-56079) as well as, separately, the night where an elevated VHE state was detected (MJD 56067).   This flare is excluded from the dark run analysis results.  The results are shown with statistical errors only in Table 1.   The systematic error on the energy scale is estimated between 20 and 35\%.

The hour-scale exposures during the first dark run show the blazar to be at an average flux of (8.6$\pm$3.6) $\times10^{-8}$ ph m$^{-2}$s$^{-1}$ above the observational energy threshold of 315 GeV (approximately 8\% of the Crab Nebula flux above this same threshold).    The first dark-run observations are paired with two contemporaneous \swift exposures, described in Section 2.3.  

On MJD 56067, VERITAS detected a short-lived VHE flare of \es.  These observations show the blazar flux to rise from $\sim$50\% to 120\% of the Crab Nebula in less than 30 minutes (see top panel of Figure 3).  The rise in flux was immediately followed by a decay, dropping back to $\sim40\%$ of the Crab Nebula approximately 90 minutes after the start of the event.   The probability that the blazar flux on MJD 56067 was constant is less than 0.2\% (see Figure ~3, $\chi^2=29.5$ with 11 degrees of freedom).  The first 0.7 ks (12 minutes) of the VERITAS observations of the flaring event on MJD 56067 were matched with simultaneous \swift observations, described in Section 2.3.  VERITAS continued to observe \es\, through 1 June 2012, detecting the source at an average of 12\% Crab Nebula flux (1.5$\pm$0.2 $\times10^{-7}$ ph m$^{-2}$s$^{-1}$ above 315 GeV). 

\subsection{\textit{Fermi} Large Area Telescope}
The \fermi Large Area Telescope (LAT) is a space-based telescope that typically monitors the entire high-energy gamma-ray sky from below 30 MeV to $\sim$300 GeV every three hours \citep{atwood}.  The instrument has better than 10\% energy resolution, with an angular resolution of better than 0.15$^{\circ}$ for energies greater than 10 GeV. 

Spectral analysis was completed for the period between (MJD 56054 to 56082) with the intention to search for an elevated gamma-ray state occurring contemporaneously with the elevated VHE states observed on MJD 56062 and MJD 56067.  All analysis was completed with  \texttt{FermiTools v9r27p1}.   Events were extracted from a 30$^{\circ}$-radius region centered on the \es\, coordinates.   ``Diffuse class" events with zenith angle of $<100^{\circ}$ and energy between 300 MeV and 300 GeV were selected.  In order to reduce contamination from Earth-limb gamma rays, only data taken while the rocking angle of the satellite was less than 52$^{\circ}$ were used.  The significance and spectral parameters were calculated using the unbinned maximum-likelihood method \textit{gtlike} with the \texttt{P7SOURCE\_V6} instrument-response functions.  The background model was constructed to include nearby ($<30^{\circ}$ away) gamma-ray sources from the second \fermi LAT catalog (2FGL, \citealt{nolan}) as well as diffuse emission.

As in the 2FGL catalog, a log-parabolic function was used for nearby sources with significant spectral curvature and a power law for those sources without spectral curvature.  The spectral parameters of sources within 7$^{\circ}$ of \es\, were left free during fitting, while those outside of this range were held fixed to the 2FGL catalog values.  The Galactic diffuse emission was modeled with the file \texttt {gal\_2yearp7v6\_v0.fits} and the isotropic emission component was modeled with the file \texttt{iso\_p7v6source.txt}.\footnote{Available at \tt http://fermi.gsfc.nasa.gov/ssc/data/analysis/software/aux.}

The analysis of LAT data between MJD 56054 and 56082 results in a test statistic (TS; \citealt{mattox}) of 97 above 100 MeV.  The spectral fitting shows the source to be in a slightly elevated state as compared to the 2FGL value (less than double the 2FGL integral flux of F$_{1-100{\rm GeV}}=(8.8\pm0.3)\times10^{-9}$ ph cm$^{-2}$s$^{-1}$), with an integral flux of (1.6$\pm0.8)\times10^{-8}$ ph cm$^{-2}$s$^{-1}$ above 1 GeV and a spectral index of 1.9$\pm0.1$.  

The data were binned in time to search for evidence of an elevated gamma-ray flux (see Figure 4).  No evidence for variability on weekly timescales is found; the data are consistent with a steady flux at the 99.75\% confidence level  ($\chi^2$=0.648 for 3 degrees of freedom).  The LAT observations do not provide sufficient statistics for a more detailed investigation of variability, i.e. on daily timescales, which result in time-bin TS values of less than 9.    Upper limits at 95\% confidence are derived for daily LAT exposures of 1ES\,1959+650 and are displayed in Figure 4 by downward pointing arrows.

\subsection{Swift XRT}
The \swift X-ray Telescope (XRT) is a space-based grazing-incidence Wolter-I telescope which focuses X-rays between 0.2 keV and 10 keV onto a 110 cm$^2$ CCD \citep{gehrels}.  The \swift telescope took 16 windowed timing (WT) exposures of \es\, between 19 April 2012 and 1 June 2012 (MJD 56036 - 56079), each between 0.5 ks and 1.5 ks long.  The exposure on MJD 56067 is strictly simultaneous with VERITAS observations.

The data were analyzed using the \texttt{HEASoft} package Version 6.12.  Rectangular source regions of length and width 45 pixels and 8 pixels, respectively, were used.  Similarly sized regions of nearby source-free sky were used to estimate the background.  The exposures were grouped to ensure a minimum of 20 counts per bin and were fitted with an absorbed power-law model of the form $F(E)_{PL}= Ke^{-N_{\rm HI}\sigma(E)} (E/1$keV$)^{-\alpha}$, with a free neutral hydrogen column density parameter $N_{\rm HI}$, as in \cite{furniss}, to allow for additional absorption of soft X-rays by intervening gas in the vicinity of the blazar or along the line of sight.  The model also contains a fitted normalization parameter $K$ and the non-Thompson energy-dependent photoelectric cross section $\sigma(E)$, as taken from \cite{xraycrosssection}.

The flux and indices derived from the \swift XRT observations of \es\, are shown in the second and third panels from the top in Figure 2 and are summarized in Table 2.  The fitted $N_{\rm HI}$ was consistently $\sim$2 times higher than the Galactic value of 1$\times10^{21}$cm$^{-2}$, as measured by the LAB survey \citep{kalberla}.  The integral 2-10 keV flux recorded by the XRT ranged between 4.2  $\times10^{-11}$ erg cm$^{-2}$s$^{-1}$ and 12.9 $\times10^{-11}$ erg cm$^{-2}$s$^{-1}$ with an average flux of 7.5$\times10^{-11}$ erg cm$^{-2}$s$^{-1}$, with photon indices ranging from $\alpha=2.5$ to 3.1.  

The X-ray emission displayed by \es\, is relatively steady in the first five exposures.  The exposure on 20 May 2012 (MJD 56067) is the only strictly simultaneous \textit{Swift} observation with VERITAS, overlapping with the first 0.7 ks ($\sim12$ minutes) of VERITAS observations during the VHE flare from \es.  The steady 0.3 - 10 keV count rate observed by XRT (as shown in the bottom panel of Figure 3) shows that at least the first 12 minutes of the VHE flaring episode is not matched with a simultaneous elevated X-ray state.  

The \textit{Swift}-XRT observations on MJD 56074 show the blazar to be in an elevated state with a flux level of $(12.9\pm0.1)\times10^{-11}$ erg cm$^{-2}$s$^{-1}$, nearly twice the average of 7.5$\times10^{-11}$ erg cm$^{-2}$s$^{-1}$.  This high state is observed to drop to approximately half the average (4.0 $\times10^{-11}$ erg cm$^{-2}$s$^{-1}$) in less than two days. No contemporaneous high state is observed in the VERITAS data, but no firm conclusions can be drawn due to the non-simultaneous nature of the exposures.

\subsection{UV and Optical}
\textbf{Swift UVOT}
The \textit{Swift}-XRT observations were supplemented with simultaneous UVOT exposures taken in the UVW1, UVM2, and UVW2 bands \citep{poole}.  The UVOT photometry is performed using the \texttt{HEASoft} program \textit{uvotsource}.  The source region consists of a single circle with $5\arcsec$ radius, while the background region consists of several 15$\arcsec$-radius circles of the nearby sky lacking visible sources. The results are corrected for reddening using E(B-V) coefficients from \cite{schlegel}, with the Galactic extinction coefficients applied according to \cite{fitzpatrick}.  The largest source of error derived for the intrinsic flux points is due to the uncertainty in the reddening coefficients E(B-V).  The UVOT W1, W2 and M2 flux values derived from the observations are shown in Figure 2 (fourth panel from the top), and summarized in detail in Table 3.  These exposures show relatively steady flux values, with a small decrease of $\sim30\%$ up to MJD 56074, the day where an elevated X-ray state was observed with XRT.

\textbf{Super-LOTIS} The Super-LOTIS robotic 0.6-meter telescope located on Kitt Peak in Arizona took R-band exposures of \es\, between MJD 56034 and MJD 56080.  During each night, three individual frames were acquired with the standard Johnson-Cousins R-band filter. Each image was reduced using an analysis pipeline that, after subtracting the bias and the dark current, combines the flat-fielded frames in a single image for each night. Aperture photometry with a circular aperture of 15$\arcsec$ was performed for both the blazar and each of the seven reference stars detailed in the Landessternwarte Heidelberg-K\"{o}nigstuhl catalogue\footnote{\tt http://www.lsw.uni-heidelberg.de/projects/extragalactic/charts/} with a circular aperture of 15$\arcsec$. This aperture is large enough to encompass all the light enclosed in the irregular PSF. The local sky level is computed in a circular annulus of inner/outer radius of 18$\arcsec$/25$\arcsec$. The final flux values for \es\, are calculated by applying the photometric zero-point derived for each night, comparing the instrumental magnitude of the reference stars to the known magnitudes in the R-band.

The R-band monitoring data from Super-LOTIS are shown in the bottom panel of Figure 2 and are summarized in Table 4.  The observations show a relatively steady optical magnitude of between 13.8 and 14.0, with a conservative photometry error estimate of $\pm$0.1 optical magnitude.

\textbf{iTelescope} V-band and R-band exposures were taken by iTelescope between MJD 56028 and 56080.  iTelescope is a robotic telescope system located in Nerpio, Spain \footnote{\tt http://www.itelescope.net/}. The telescopes used are twins (T07 and T18), and are each of 431\,mm (17-inch) aperture at $f$/6.8. They employ an SBIG STL-1100M CCD camera.  The V filter is a standard Johnson-Cousins set.  The R filter is not standard and requires a color correction, where the addition of approximately 0.040 optical magnitude transfers the non-standard filter magnitudes to the standard Johnson-Cousins R.  The data were reduced with \texttt{MIRA Pro Version 7.0}. The reduction is with standard aperture (radius of 5$\arcsec$) photometry, using the same standard stars as were used for the Super-LOTIS data reduction. 

The V-band and R-band data are shown in the bottom panel of Figure 2 and summarized in Table 4.  These observations show elevated luminosity in both the V and R bands (by approximately 0.2 optical magnitude) on two of the days where Super-LOTIS also provides R-band measurements approximately 7.5 hours after the iTelescope exposures were taken (shown in bold in Table 4).  Comparison with the contemporaneous Super-LOTIS R-band measurements suggests that the blazar exhibits a small level of intranight variability.  This fast variability occurs on the same night as the X-ray flux is observed to be high.

\subsection{Summary of Observations}
The broadband observations summarized above show a VHE flare on MJD 56067, where no elevated X-ray state is observed simultaneously for the first 12 minutes of the VHE flaring event.  Additional variability is observed over the full window of observation, including an X-ray flux increase and intranight X-ray variability on MJD 56075.  The X-ray flux was observed to drop over the next two days, with no corresponding (non-simultaneous) change in VHE flux observed.  

\section{Modeling}

\subsection{Time Independent Description (SSC)}
We apply a time-independent SSC model to the relatively low and elevated flux states of \es\, observed on MJD 56064 and 56067.  Both of these days have sufficient multiwavelength coverage to provide a full view of the broadband spectral energy distribution.  Since the VHE flux of \es\, observed on MJD 56064 is consistent with the average flux from the two dark runs (excluding the VHE state on MJD 56067), the VHE spectrum is represented by the spectrum averaged over the two dark runs.  Moreover, the average LAT spectrum is used to represent both the low and high state since no significant variability was detected within the LAT energy band. 

The SSC model applied to the data is described in detail in \cite{markus}.   This model is an equilibrium SSC model with emission originating from a spherical region of relativistic electrons with radius $R$.  This region moves down the jet with a Lorentz factor $\Gamma$.  The jet is oriented such that the angle with respect to the line of sight is $\theta_{\rm obs}$. In order to minimize the number of free parameters, the modeling is completed with $\theta_{obs} = 1/\Gamma$, for which $\Gamma = D$, where $D$ represents the Doppler boosting factor.

Electrons are injected into the spherical region with a power-law distribution of $Q(\gamma) = Q_0 \gamma^{-q}$ between the low- and high-energy cut-offs, $\gamma_{min,max}$. The electron distribution spectral indices used for 1ES\,1959+650 are $q = 2.7-2.8$, which can be produced under acceleration in relativistic oblique shocks \citep{summerlin}.  In order to reach an equilibrium state, the model evaluates the steady state produced when considering particle injection, radiative cooling and particle escape.  The particle escape is characterized with an efficiency factor $\eta$, such that the escape timescale $t_{\rm esc} = \eta \, R/c$, with $\eta=1000$ for this work, setting up an equilibrium scenario with a relatively long escape timescale for the relativistic particles.  The variability timescale $t_{var}$ is determined by the light crossing timescale of the emitting region ($t_{var}=\delta R/c$).   According to this SSC model, the particle distribution streams along the jet with a kinetic power $L_e$.  Synchrotron emission results from the presence of a tangled magnetic field $B$, with a Poynting flux luminosity of $L_B$.  The parameters $L_e$ and $L_B$ allow the calculation of the equipartition parameter $L_B/L_e$.  

A reasonable description to the low state (MJD 56064) is achieved with the parameters summarized in Table 5, and displayed by the solid line in Figure 5.  Starting from this low state representation, two possible scenarios are explored to describe the elevated VHE state observed on MJD 56067.   In the first realization (Scenario I, dotted line in Figure 5), the gamma-ray peak was shifted to a slightly higher energy by increasing the low-energy cut-off ($\gamma_{min}$). In order to keep the synchrotron peak at the same energy, the magnetic field is lowered.  The lower magnetic field results in a lower synchrotron power, requiring an increase in overall jet power ($L_e$) to achieve the same synchrotron luminosity.  This emission scenario, however, over-shoots the gamma-ray flux unless the radius of the emission region $R$ is increased, lowering the compactness of the emission zone.  The result of these parameter changes are shown by the dotted line in Figure 5, and summarized in the second column for Table 5.  Under this representation, the variability timescale changes from 8.2 hours in the low state to 31 hours in Scenario I.  This timescale is longer than the observed variability timescale on MJD 56067, where the VHE flux was observed to increase from 0.6$\times10^{-6}$ ph cm$^{-2}$s$^{-1}$ to 1.4$\times10^{-6}$ ph cm$^{-2}$s$^{-1}$ above 315 GeV in less than one hour.  

The eight parameters (including $L_e$) which describe the SSC emission scenario are known to be degenerate.  As a result, the parameter changes described in Scenario I are not the only changes which will account for the difference between the SEDs observed on MJD 56064 and MJD 56067.  Alternatively, in addition to the changes to the $\gamma_{min}$ and magnetic field, an increase of the Doppler factor instead of a change to the emission region size can provide a similar result (Scenario II, dashed line in Figure 5, third column of Table 5).  With this scenario, the variability timescale is still left to be relatively short (5.1 hr), in agreement with the fast flux variability observed in the VHE band on MJD 56067.  %In each case, a viewing angle of $\theta_{obs}=1/\Gamma$ ($\Gamma=D$) is used.

As seen in Figure 5, both of the elevated state SSC scenarios predict an approximate doubling of the high-energy gamma-ray flux.  This type of variability is impossible to rule out without detection of 1ES\,1959+650 by the \textit{Fermi} LAT on day timescales.  The daily 95\% confidence level upper limit on MJD 56067 (Figure 4) is more than double the LAT flux for the entire period (also used to derive the spectrum shown in the SED), indicating that the SSC-inferred LAT flux in an elevated state is still consistent with the observations.   Therefore, the broadband SEDs of 1ES\,1959+650 on MJD 56064 and MJD 56067 can be represented by a SSC emission model, necessitating multiple parameter changes from high states relative to low states in order to produce an elevated VHE state with no change to the synchrotron peak.

\subsection{Time Dependent Description (Reflected-Emission)}
In this section we present a possible scenario to describe the VHE variability detected during the contemporaneous multiwavelength observations of 1ES 1959+650. The main emphasis of our discussion is to show that the scenario can explain the data.  Our choice in model is motivated, in part, by the evidence for intervening gas within the blazar.  We apply a similar reflected-emission model to that which was used to describe the ``orphan" flaring activity of \es\, \citep{boettcher05}.  This model follows X-ray emission from a newly ejected component (blob) in the jet as it is reflected off dilute gas and/or dust in the vicinity of the jet.  The reflected emission then re-enters the jet before the blob, which is moving down the jet, reaches the location of the reflector.  The application of this model is notably distinct from that applied in \cite{boettcher05}, where for this application the incident flux on the cloud is integrated over the time it takes the blob to pass the reflecting cloud instead of taken from a single short-lived X-ray flaring period.  This integration over the blob's travel is necessary due to the assumed parsec-scale proximity of the blob to the reflecting cloud.

We assume that at a distance $R_m = 1 \, R_{\rm m, pc}$~pc from the central engine, 
moderately dense clouds of gas and dust (hereafter referred to as the ``mirror'')
intercept the synchrotron emission from portions of the jet located inside $R_m$, 
and reprocess part of this flux back into the jet trajectory. Following \cite{boettcher05}, 
the distance $R_m$ can be related to the observed time delay between the emergence 
of a new jet component from the core, and the (observer's frame) time at which the 
new component intercepts the mirror. For a characteristic bulk Lorentz factor of the
new component of $\Gamma = 10 \, \Gamma_1$, and a time delay of, for example, $\Delta t \sim 5 \times 10^5$~s:

\begin{equation}
\Delta t \sim {R_m \over 2 \, \Gamma^2 \, c} \sim 5 \times 10^5 \, R_{\rm m, pc} \, 
\Gamma_1^{-2} \; {\rm s}
\label{Rm},
\end{equation}
according to which the new component would have emerged around MJD 56062 for $\Gamma \sim 10$
and $R_m \sim 1$~pc. 

%Using an X-ray flux of order three times the average 2-10 keV flux 
%measured by the XRT ($F_x^{\rm flare} \sim 3 \times 10^{-10}$~erg~cm$^{-2}$~s$^{-1}$), 
%we can estimate the corresponding flux at the location of the reflecting cloud, as

%\begin{equation}
%F_m \sim {d_L^2 \over R_m^2} F_x^{\rm flare} \sim 1.1 \times 10^7 \; \Gamma_1^{-4}
%\; {\rm erg \; cm}^{-2} \; {\rm s}^{-1},
%\label{Fmflare}
%\end{equation}
%where $d_L$ is the luminosity distance to earth.

The accumulated and reprocessed jet-synchrotron flux will be intercepted by the blob within 
the time interval between emitting the first photons at the time of emergence of the new
component, and intercepting the location of the cloud. This time is also characteristic 
of the time during which the cloud receives this flux, and can be estimated as is done 
in \cite{boettcher05}

\begin{equation}
\Delta t_{\rm fl} \sim {R_m \over 8 \, \Gamma^4 \, c} \sim 1.3 \times 10^3 \,
R_{\rm m, pc} \, \Gamma_1^{-4} \; {\rm s}.
\label{tfl}
\end{equation}
With this time scale, we can calculate an average flux received by the cloud as

\begin{equation}
F_m^{\rm ave} \approx {F_x^{\rm qu} \, d_L^2 \over c \, \Delta t_{\rm fl}}
\int\limits_{R_b}^{R_m} {dx \over x^2} \approx {F_x^{\rm qu} \, d_L^2 \over 
c \, \Delta t_{\rm fl} \, R_b} \sim 5.3 \times 10^{13} \; R_{\rm m, pc}^{-1} 
\, \Gamma_1^4 \, R_{b, 16}^{-1} \; {\rm erg \; cm}^{-2} \; {\rm s}^{-1},
\end{equation}
\label{Faverage}
where we have used an estimate of the quiescent synchrotron X-ray flux from \es\, 
of $F_x^{\rm qu} = 1.0 \times 10^{-10}$~erg~cm$^{-2}$~s$^{-1}$ (similar to the 
average found from the \swift XRT observations), and $R_b = 10^{16}\, R_{b, 16}$~cm 
is the radius of the newly emerged jet component (the ``blob''). For a characteristic 
density of the clouds of $n_c \sim 10^6$~cm$^{-3}$, this leads to an ionization 
parameter of

\begin{equation}
\xi_{\rm ion} \equiv {F_m^{\rm ave} \over n_c} \sim 5 \times 10^7 \; {\rm erg \; cm
\; s}^{-1}.
\label{xi_ion}
\end{equation}

\noindent This implies that any dust is expected to be destroyed, and all gas to 
be highly ionized by the impinging X-ray emission.  If the cloud is thermally 
reprocessing this flux, this would require an equilibrium temperature $T_{\rm equi} 
\sim 36000$~K, which also requires the gas to be highly ionized.  We therefore 
conclude that the most likely mode of reprocessing the accumulated jet synchrotron 
emission is Compton reflection off free electrons in the highly ionized cloud.   It is not necessary that this is the only cloud within the vicinity of the blazar, maintaining the possibility that there is additional neutral absorbing gas surrounding the blazar, as found in \cite{furniss}.  We only assume that a blob moving relativistically down the jet passes sufficiently close to a cloud that radiation from the blob can temporarily ionize the cloud, so that for a short time the blob emission is reprocessed via Compton reflection off free electrons in the cloud.  This ionized gas might also act as a shield to molecular gas such as CO, as 
predicted in photodissociation region models (e.g., \citealt{tielenn,krumholz,gloverclark}).

The characteristic photon frequency of jet synchrotron photons from 1ES\,1959+650 
is $\nu_{\rm sy} \sim 10^{17}$~Hz (see Figure 5), corresponding to a normalized photon 
energy of $\epsilon_{\rm sy} \equiv h \nu_{\rm sy} / (m_e c^2) \sim 10^{-2}$. Upon 
Compton reflection by the cloud, this will be boosted in the jet rest frame to 
$\epsilon'_{\rm sy} \sim 0.1 \, \Gamma_1$. Therefore, any relativistic electrons 
(with $\gamma_e \gtrsim 10$) will interact with these reflected photons in the 
Klein-Nishina regime, resulting in strongly suppressed Compton scattering, making 
the production of a gamma-ray flare after the emergence of a new blob within the jet
unlikely in a purely leptonic scenario.

It can be seen that the bulk Lorentz factor of the blob is a critical unknown parameter 
in this model, with key derived parameters strongly depending on it. The observed 
synchrotron peak frequency of \es\, is $\sim1\times10^{17}$ Hz. Therefore, even 
without any blue-shifting from bulk motion, Compton scattering in the Thomson
regime happens up to $\epsilon_{\rm C}\sim1/\epsilon_{\rm sy}\sim100$, corresponding 
to $\sim 50$~MeV. Therefore, even for $\Gamma = 1$, a synchrotron mirror scenario 
would not efficiently produce VHE $\gamma$-rays via Compton scattering.

Relativistic protons with Lorentz factors of $\gamma_p \gtrsim 6 \times 10^3$, on 
the other hand, can interact with the reflected photons through pion production 
processes. Following the analysis in \cite{boettcher05} and using the average flux 
$F_m^{\rm ave}$, we find that producing a VHE 
$\gamma$-ray flare with a luminosity of $L_{\rm VHE} \sim 1.5 \times 10^{45}$~erg~s$^{-1}$ 
requires a total number density $n_p$ of relativistic protons

\begin{equation}
n_p \sim 1.4 \times 10^5 \, R_{\rm m, pc} \, \Gamma_1^{-12} \, \tau_{-1}^{-1} \, R_{b, 16}^{-2}
\; {\rm cm}^{-3},
\label{Np}
\end{equation}

\noindent where $\tau_m = 0.1 \, \tau_{-1}$ is the fraction of incident flux reflected 
by the cloud.  We have assumed that the protons have a power-law distribution in energy 
$N(\gamma_p) \propto \gamma_p^{-2}$ with a low-energy cutoff at  $\gamma_{\rm p, min} = 
\Gamma$.  This corresponds to a total (co-moving  frame) energy in relativistic protons 
in the blob of

\begin{equation}
E'_{b, p} \sim 3.2 \times 10^{49} \, R_{\rm m, pc} \, \Gamma_1^{-11} \, \tau_{-1}^{-1} \, R_{b, 16}
\; {\rm erg},
\label{Ep}
\end{equation}
and a kinetic power in the jet, carried by relativistic protons, of

\begin{equation}
L_p \sim 7.3 \times 10^{45} \, R_{\rm m, pc} \, \Gamma_1^{-9} \, \tau_{-1}^{-1} \, f \;
{\rm erg \; s}^{-1},
\label{Lp}
\end{equation}
where $f$ is the filling factor of the jet, i.e., the fraction of the jet length occupied by 
plasma containing relativistic protons. 

The power requirement in Eq. \ref{Lp} is quite moderate if one allows for a plausible filling 
factor $f \lesssim 0.1$. Also, note the extremely strong dependence of the estimates in 
Eqs. \ref{Np} -- \ref{Lp} on the Lorentz factor. A value of $\Gamma$ just slightly above 
10 will reduce all energy requirements to very reasonable values, corresponding to a population 
of relativistic protons with energies between 10 and 100 TeV.

\section{Discussion and Conclusions}
We report contemporaneous broadband observations of the VHE-emitting blazar \es, including 0.7 ks of strictly simultaneous \textit{Swift} and VERITAS observations occurring during a period of elevated VHE flux. This blazar has shown extreme flaring episodes with uncorrelated variation in the synchrotron and gamma-ray SED components in the past \citep{krawczynski}, which could be described in a reflected non-thermal emission environment, with a blob of relativistic particles moving toward a dilute reflector made of gas or dust intrinsic to the blazar \citep{boettcher05}. 

The application of an equilibrium SSC model to the relatively low and high states of 1ES\,1959+650 on MJD 56064 and MJD 56067 is possible, with multiple parameter changes required for the synchrotron peak to remain unchanged during the elevated gamma-ray state.  Two scenarios provide a reasonable representation of the observed VHE elevated state.  One realization utilizes an increase in both the emission region size and low-energy cutoff while at the same time a decrease in the magnetic field.  The second scenario is derived from increasing the Doppler factor instead of changing the emission region size.  Both of these scenarios also predict an increase in the high-energy gamma-ray flux, which is not ruled out by the \textit{Fermi} LAT daily upper limits.  The second scenario (Scenario II) is preferred due to the hour-scale flux variability that is maintained with the parameter changes.

Motivated by the possibility of uncorrelated variability as well as compelling evidence for the existence of dilute gas in the vicinity of the blazar, we investigate these broadband observations using the reflected-emission paradigm.   We find that the resulting ionization of the cloud and dust makes Compton reflection on free electrons the most likely mode of reprocessing the jet synchrotron emission.  The emission from the ionized reflector re-enters the blob in the Klein-Nishina regime, suppressing leptonic Compton upscattering that might be responsible for an elevated gamma-ray state with no corresponding increase in X-ray state.

The production of an elevated gamma-ray component, however, is still possible if there are hadrons within the blob with energies greater than 10 TeV.  This hadronic synchrotron reflection model, in which relativistic protons interact with the reflected emission to produce charged and neutral pions, provides a possible explanation of the uncorrelated gamma-ray variability as inferred from the broadband observations.  

Evidence for hadrons as the source of the highest-energy emission from blazars would highlight these galaxies as possible progenitors of cosmic rays.  However, the energy of the hadrons predicted by this model peak around 10 TeV, which is insufficient as an explanation for the source of the ultra-high-energy cosmic rays. Detection of a neutrino flux in the direction of \es\, would provide compelling evidence that the observed non-thermal emission is derived from hadronic interactions.  However, the expected neutrino number flux for the outlined scenario would be within a factor of two of the VHE-photon number flux and is too low to be detected by current-generation neutrino detectors such as IceCube \citep{reimer}.  Stronger conclusions regarding the non-thermal emission mechanism at work within the jet of \es\, and a more reliable application of time-dependent model such as a reflected emission model would be possible with a more comprehensive broadband dataset including high-cadence simultaneous observations.

\acknowledgments
This research is supported by grants from the U.S. Department of Energy Office of Science, the U.S. National Science Foundation and the Smithsonian Institution, by NSERC in Canada, by Science Foundation Ireland (SFI 10/RFP/AST2748) and by STFC in the U.K. We acknowledge the excellent work of the technical support staff at the Fred Lawrence Whipple Observatory and at the collaborating institutions in the construction and operation of the instrument.  The VERITAS Collaboration is grateful to Trevor Weekes for his seminal contributions and leadership in the field of VHE gamma-ray astrophysics, which made this study possible.

{\it Facilities:} \facility{VERITAS}, \facility{Fermi}, \facility{Swift}, \facility{Super-LOTIS}, \facility{iTelescope}.

\clearpage

%Figures

\begin{figure}
\epsscale{1.1}
\plotone{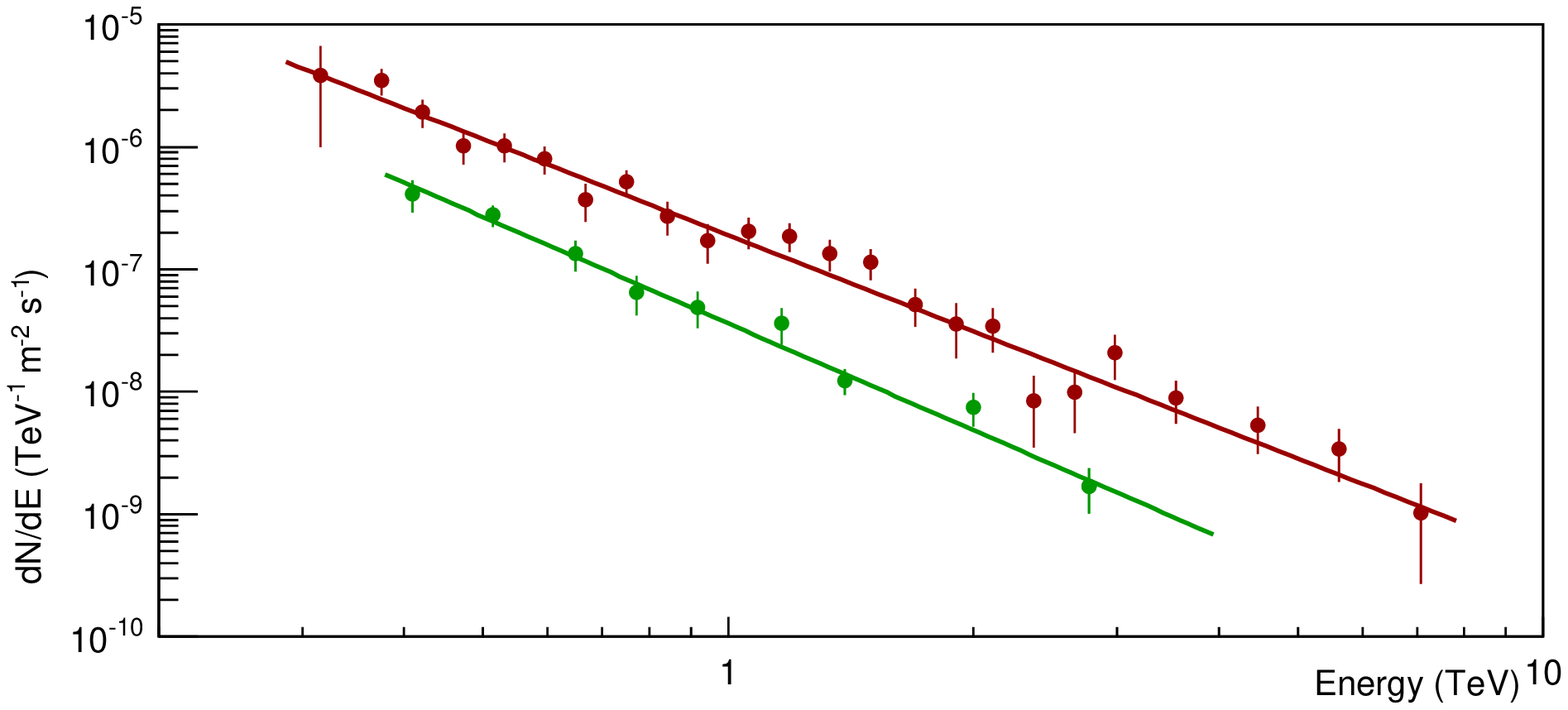}
\caption{The VERITAS-measured spectra of \es\, averaged over both dark-runs and excluding the flaring period (MJD 56034-56079; green) and during the flare (MJD 56067; red).  The spectrum measured over both dark-runs (without the data from 56067) is used for the low state of the SSC modeling.  The spectra are shown with 1$\sigma$ statistical errors.  The power-law fitting results are summarized in Table~1.
\label{fig1}}
\end{figure}

\begin{figure}
\epsscale{1.1}
\plotone{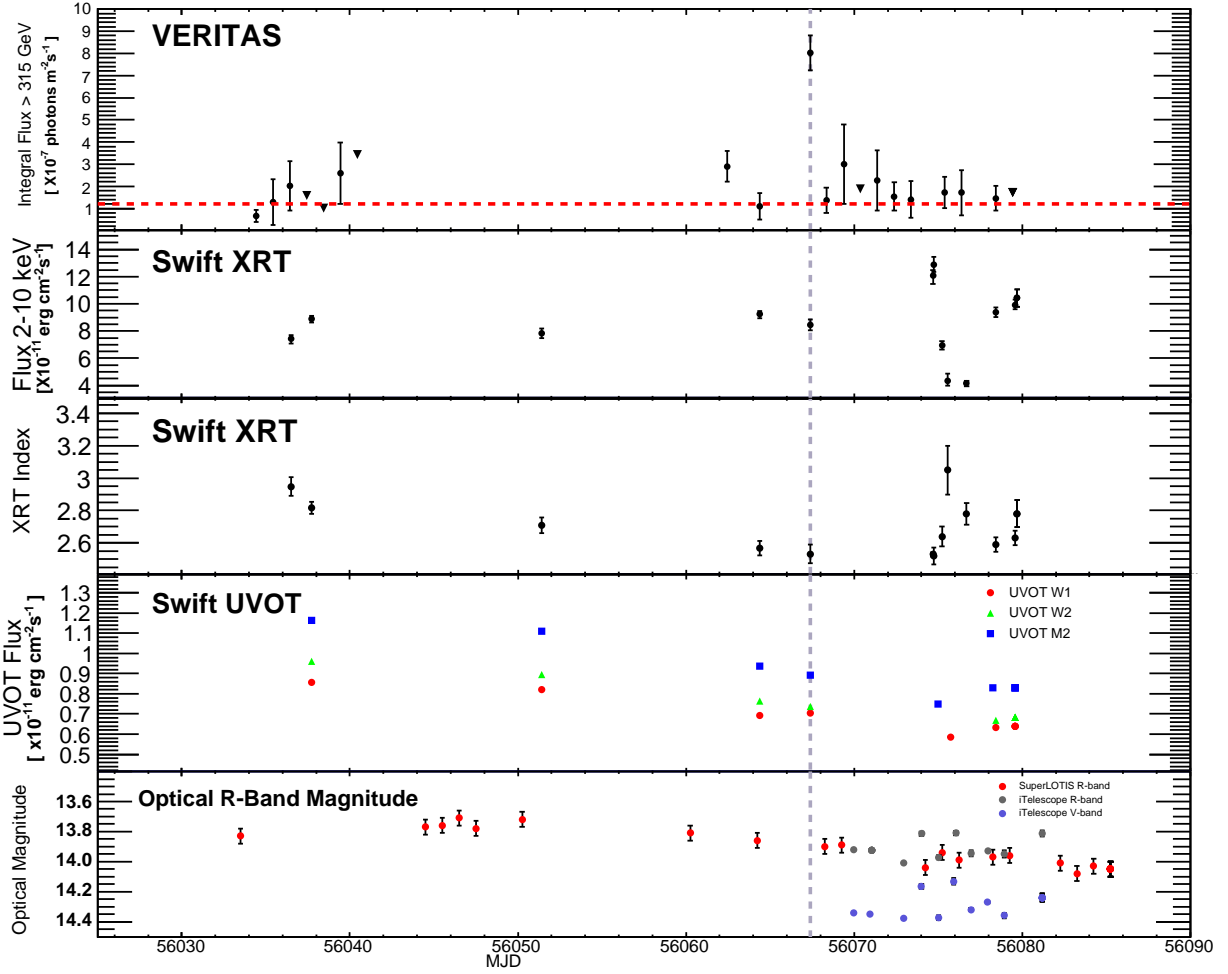}
\caption{Broadband observations of \es\, in April and May of 2012.  The top panel shows VERITAS integral flux values above 315 GeV, denoting 95\% confidence upper limits with downward pointing arrows.  The flux points are shown with 1$\sigma$ statistical error bars.  A line denoting 10\% Crab Nebula above the same threshold is denoted by the red dotted line.  The day of the VHE flare (MJD 56067) is denoted by a grey dotted line. In the two panels below this, the \textit{Swift} XRT flux and spectral indices are shown.  During the flare, VERITAS observed a maximum flux of 10 times the average flux of the darkrun, lasting less than two hours, with no change observed simultaneously in the X-ray flux as observed by the \textit{Swift} XRT.   The W1, W2 and M2 bands from UVOT exposures similarly show no evidence of increased UV flux during the VHE flare.  In the bottom panel are shown observations in the R and V bands from the Super-LOTIS and iTelescope. 
\label{fig2}}
\end{figure}

\begin{figure} 
\epsscale{1.0}
\plotone{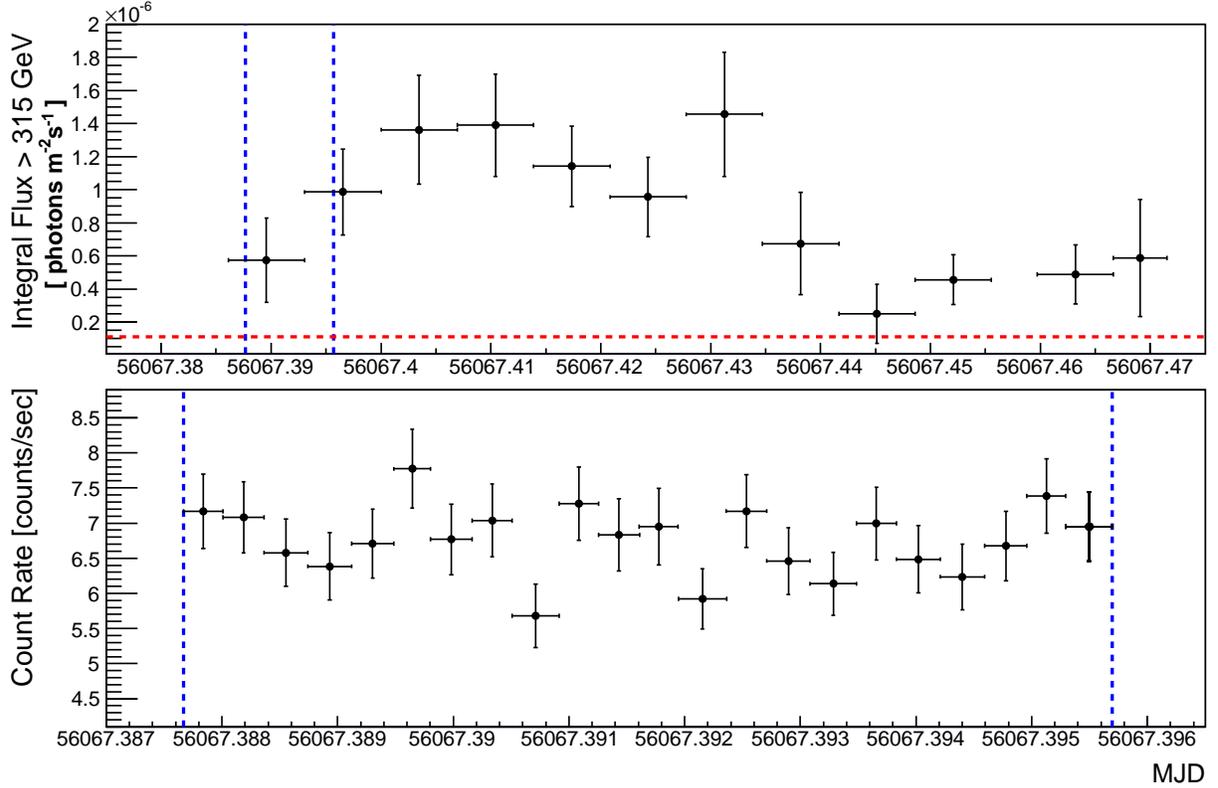}
\caption{In the top panel, the VHE flux of \es\, above 315 GeV as observed by VERITAS on MJD 56067 in ten-minute time bins.  The flux points are with with 1$\sigma$ statistical error bars.  The red-dotted line represents 10\% Crab Nebula flux above the same threshold.  The start and end of the simultaneous \textit{Swift} observations are denoted with dotted blue lines (spanning 12 minutes).  In the bottom panel, the \textit{Swift} XRT 0.3-10 keV count rate is shown over the simultaneous observation interval.
\label{fig2}}
\end{figure}

\begin{figure}
\epsscale{0.9}
\plotone{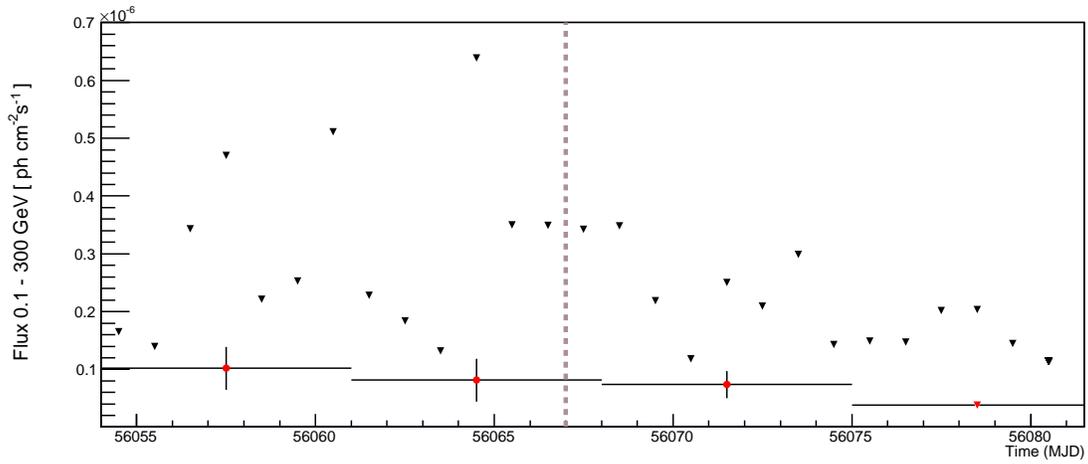}
\caption{The high energy gamma-ray light curve of 1ES\,1959+650 as observed by \textit{Fermi} LAT.  Analysis is completed over four weeks on both a daily and weekly basis from MJD 56054 through MJD 56082.  Upper limits are calculated from epochs where the TS value is less than 9, which was every day in the four week window, as well as the bin corresponding to the last week.  There is no indication of detected high-energy gamma-ray variability during this time period.  The time corresponding to the elevated state of the VHE data is highlighted by the grey dotted line.
\label{fig2}}
\end{figure}

\begin{figure} 
\epsscale{1.0}
\plotone{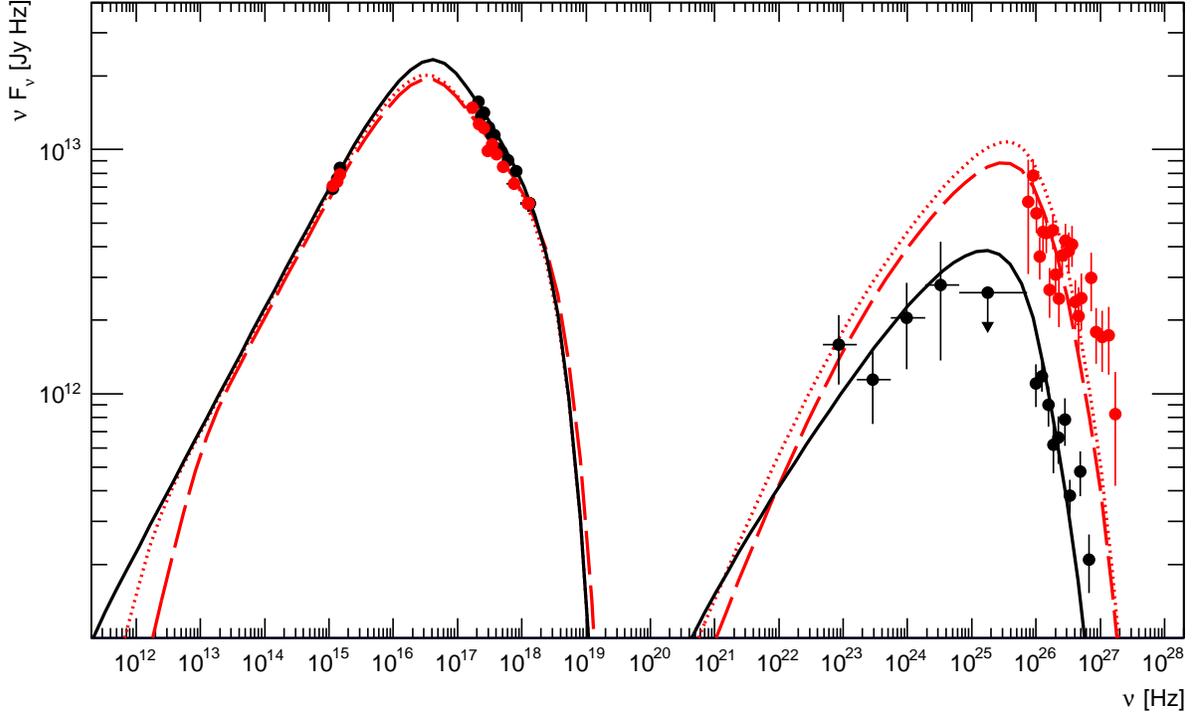}
\caption{Broadband SED of 1ES 1959+650 with data from MJD 56064 (black) and MJD 56067 (red).   The VHE spectrum for the low state is represented by the average spectrum measured over the two dark runs, excluding the state on MJD 56067. These data are explored with a SSC representation, where the black line corresponds to the low state and the red dashed and dotted lines correspond to the high gamma-ray state observed on MJD 56067.  The dotted line is produced by increasing the emission region size and low-energy cutoff, while the magnetic field is decreased.  The dashed line representation is obtained by increasing the Doppler factor and keeping the emission region size constant, in addition to increasing the low energy cutoff and decreasing the magnetic field.  All parameter values used in the modeling of each state are summarized in Figure 5.
\label{fig4}}
\end{figure}

%Tables
\begin{deluxetable}{lcccccccccc}
\tabletypesize{\scriptsize}
\rotate
\tablecaption{Summary of VERITAS observations and spectral fits, shown with statistical (1$\sigma$) errors.  The April, May and combined dark runs do not include the flare period.  The combined dark runs and flare are shown in Figure 1 with green and red lines, respectively, showing the power-law fits to each dataset. \label{tbl-1}}
\tablewidth{-10pt}
\tablehead{
\colhead{} &\colhead{Exposure} & \colhead{Exposure} & \colhead{Detection} & \colhead{Number} & \colhead{Number} & \colhead{Number}  &\colhead{Power law} & \colhead{Average Integral} &\colhead{Integral Flux} & \colhead{Power-Law} \\
\colhead{} & \colhead{Date} & \colhead{Livetime} & \colhead{Significance} & \colhead{ON} & \colhead{OFF} & \colhead{Excess}   &\colhead{Index} & \colhead{Flux $\ge$ 315 GeV} &\colhead{Percent Crab}& \colhead{Fit} \\
\colhead{} &\colhead{MJD} & \colhead{[minutes]} & \colhead{$\sigma$} & \colhead{Events} & \colhead{Events} &\colhead{Events} &\colhead{$\Gamma$} & \colhead{$\times10^{-7}$ [ph m$^{-2}$s$^{-1}$]} &\colhead{[\%]}& \colhead{$\chi^2$/DOF} \\ 
}
\startdata
April Dark Run &56034-56040& 186 & 6.6  &90&425&51&2.5$\pm$0.4&0.9$\pm$0.4 &  8&1.1/5 \\
%Flare on MJD 56067 &56062& 28 & 6.1  &27&60&22&3.4$\pm0.7$&2.9$\pm$0.8&24 &4.0/4 \\
 Flare &56067& 106 & 26.3  &276&249&253&2.6$\pm$0.1&8.0$\pm$0.8& 66 &19.4/22\\
  May Dark Run &56064-56079& 231 & 10.3  &151&501&106&3.2$\pm$0.3&1.5$\pm$0.2&12 &4.3/6 \\
   \hline
 Combined Dark Runs &56034-56079& 417 & 12.1  &241&926&157&3.0$\pm$0.2&1.4$\pm$0.2& 11 &3.8/7\\
\enddata
\end{deluxetable}

\begin{deluxetable}{llcccccc}
\tabletypesize{\scriptsize}
\rotate
\tablecaption{\textit{Swift} XRT summary of observations and spectral fitting results.\label{tbl-2}}
\tablewidth{0pt}
\tablehead{
\colhead{\textit{Swift} XRT} & \colhead{Observation}   & \colhead{Exposure} & \colhead{Power-law} & \colhead{ N$_{\rm HI}$}& \colhead{Integral Flux} &  \colhead{$\chi^2$\tablenotemark{\dagger}} & \colhead{Degrees} \\
\colhead{Observation}                 & \colhead{Date} & \colhead{Time}         & \colhead{Index}  & \colhead{Density} &\colhead{2-10 keV}& \colhead{} & \colhead{of} \\
\colhead{ID\tablenotemark{*}}                 & \colhead{[MJD]} & \colhead{[ks]}         & \colhead{$\alpha$}  & \colhead{[$\times10^{21}$cm$^{-2}$]} &\colhead{[$\times10^{-11}$ erg cm$^{-2}$s$^{-1}$]}& \colhead{} & \colhead{Freedom} \\
}
\startdata
00035025075     & 56036.51  &1.0  &2.95$\pm$0.06 & 2.2$\pm$0.1& 7.4$\pm$0.3&191.1 & 180     \\
00035025076     & 56037.72  &1.5    & 2.82$\pm$0.04&2.16$\pm$0.08 &8.9$\pm$0.3& 299.5&   245   \\
00035025077     & 56051.41  & 1.0 & 2.71$\pm$0.05& 1.89$\pm$0.09& 7.8$\pm$0.3&206.9&     195 \\
00035025078     &56064.38   & 1.1&2.57$\pm$0.05 &2.0$\pm$0.1 &9.3$\pm$0.3 &   225.5&  207 \\
00035025079     &56067.39   & 0.7 & 2.53$\pm$0.06 & 1.9$\pm$0.1&8.4$\pm$0.3 & 124.2&   156  \\
00035025080 (a)&56074.69   &0.6  &2.53$\pm$0.02 &2.3$\pm$0.1 &12.0$\pm$0.1 &150.0   &  170 \\
00035025080 (b)&56074.75   &0.6  &2.52$\pm$0.05 &2.2$\pm$0.1 &12.9$\pm$0.1 &  154.3  & 178 \\
00035025081 (a)& 56075.23  & 0.9 &2.64$\pm$0.06 &2.1$\pm$0.1 & 7.0$\pm$0.3& 140.9 &  153  \\
00035025081 (b)& 56075.57  &0.3  & 3.1$\pm$0.1&  2.7$\pm$0.3 &4.3$\pm$0.3& 46.4 &  46 \\
00035025082     &56076.69   & 0.9 & 2.78$\pm$0.07& 2.0$\pm$0.1& 4.2$\pm$0.3& 131.1  & 124  \\
00035025083     & 56078.43  & 1.0   &2.59$\pm$0.04 & 2.1$\pm$0.1&9.4$\pm$0.3 &  187.8 &  204 \\
00035025084 (a)& 56079.57  & 1.5   & 2.52$\pm$0.05& 1.9$\pm$0.1& 9.3$\pm$0.3& 188.4 &  171  \\
00035025084 (b)& 56079.70  & 0.5 & 2.78$\pm$0.08 &2.5$\pm$0.2 & 10.4$\pm$0.2& 101.2 &   98 \\
\enddata
\tablenotetext{*}{For observations consisting of two time separated exposures, we denote the first with (a) and the second with (b).  The MJD is given for the start of each exposure, respectively.}
\tablenotetext{\dagger}{Refers to the $\chi^2$ of the power-law fit.}
\end{deluxetable}

\begin{deluxetable}{lccc}
\tabletypesize{\scriptsize}
\tablecaption{Summary of ultraviolet observations from the \textit{Swift} UVOT\label{tbl-4}}
\tablewidth{0pt}
\tablehead{
\colhead{Date} & \colhead{UVW1}& \colhead{UVW2} & \colhead{UVM2}\\
\colhead{[MJD]} & \colhead{[$\times10^{-11}$ erg cm$^{-2}$s$^{-1}$]}& \colhead{[$\times10^{-11}$ erg cm$^{-2}$s$^{-1}$]} & \colhead{[$\times10^{-11}$ erg cm$^{-2}$s$^{-1}$]}\\
}
\startdata
 56037&8.6$\pm$0.2 &9.6$\pm$0.2 & 11.6$\pm$ 0.3\\
 56051 &8.2$\pm$0.2&8.9$\pm$0.2&   11.1$\pm$ 0.3\\
 56064 &6.9$\pm$0.2&7.6$\pm$0.2&    9.4$\pm$ 0.3\\
 56067 &7.1$\pm$0.2&7.4$\pm$0.2&   8.9$\pm$0.3\\
 56075  &5.9$\pm$0.1&--&  7.5$\pm$0.2\\
 56078  &6.3$\pm$0.2&6.7$\pm$0.2&  8.3$\pm$0.2\\
 56079    &6.4$\pm$0.2&6.8$\pm$0.2&8.3$\pm$0.2\\
\enddata
\end{deluxetable}

\begin{deluxetable}{lccc}
\tabletypesize{\scriptsize}
\tablecaption{Summary of optical observations from Super-LOTIS and iTelescope.  The two pairs of exposures in bold are taken on the same night, less than eight hours apart, and show a $\sim$0.2 magnitude difference suggesting a small level of intranight variability in the R-band.  The standard stars used to calibrate these measurements do not show any evidence of possible instrumental effects which might cause such a difference.  \label{tbl-5}}
\tablewidth{0pt}
\tablehead{
\colhead{Exposure} & \colhead{Super-LOTIS} & \colhead{iTelescope} &\colhead{iTelescope}\\
\colhead{Date} & \colhead{R-band} & \colhead{R-band} &\colhead{V-band}\\
\colhead{[MJD]} & \colhead{Magnitude} & \colhead{Magnitude} &\colhead{Magnitude}
}
\startdata
56034.0& 13.8$\pm$0.1 &&\\
56045.0& 13.8$\pm$0.1 &&\\
56046.0 &13.8$\pm$0.1&&\\
56047.0 &13.7$\pm$0.1&&\\
56048.0 &13.8$\pm$0.1 &&\\
56050.3 &13.7$\pm$0.1 &&\\
56060.3 &13.8$\pm$0.1 &&\\
56064.3 &13.9$\pm$0.1 &&\\
56068.3 &13.9$\pm$0.1 &&\\
56069.3 &13.9$\pm$0.1 &&\\
56070.0 &-- & 13.920$\pm$0.007& 14.340$\pm$0.006\\
56071.0&--& 13.926$\pm$0.003& 14.349$\pm$0.005\\
56073.0 &--&14.009$\pm$0.006& 14.378$\pm$0.007\\
\textbf{56074.0}&--& \textbf{13.814$\pm$0.014}&\textbf{14.164$\pm$0.015}\\
\textbf{56074.3} &\textbf{14.0$\pm$0.1} &&\\
56075.3 &13.9$\pm$0.1 &&\\
56075.0 &--&13.971$\pm$0.014&14.375$\pm$0.016\\
\textbf{56076.0}&--& \textbf{13.810$\pm$0.015}&\textbf{14.131$\pm$0.021} \\
\textbf{56076.3} &\textbf{14.0$\pm$0.1} &&\\
56077.0&--&13.944$\pm$0.019&14.320$\pm$0.010\\
56078.3 &14.0$\pm$0.1 &&\\
56078.0& -- & 13.928$\pm$0.007&14.270$\pm$0.005\\
56079.3 &14.0$\pm$0.1 &&\\
56079.0 &--&13.948$\pm$0.024 & 14.358$\pm$0.019\\
56081.2& --&13.814$\pm$0.020&14.239$\pm$0.027\\
56082.3 &14.0$\pm$0.1 &&\\
56083.3 &14.1$\pm$0.1 &&\\
56084.3 &14.0$\pm$0.1 &&\\
56085.3 &14.0$\pm$0.1 &&
\enddata
\end{deluxetable}

\begin{deluxetable}{lccc}
\tabletypesize{\small}
\tablecaption{Table of the SSC model parameters used for time-independent representation of the broadband data during a low state (MJD 56064) and the high state on MJD (56067).  See Section 3.1 for parameter descriptions.  The SEDs are shown in Figure 4 by a solid line for the low state, a dotted line for the high state under Scenario I and a dashed line for the high state under Scenario II.\label{tbl-6}}
\tablewidth{0pt}
\tablehead{
\colhead{SSC} & \colhead{Low State} & \colhead{High State } &\colhead{High State}\\
\colhead{Parameter} & \colhead{} & \colhead{Scenario I} &\colhead{Scenario II}\\
\colhead{[units]} & \colhead{MJD 56064} & \colhead{MJD 56067} &\colhead{MJD 56067}
}
\startdata
$\gamma_{min}$& 6$\times10^{4}$ &1.5$\times10^{5}$&9$\times10^{4}$\\
$\gamma_{max}$& 5$\times10^{5}$ &1.5$\times10^{6}$&1$\times10^{6}$\\
$q$ &2.8 &2.7 & 2.7\\
%$t_{esc}$ & 1000 & 1000& 1000 \\ % take out?
B [G] & 0.1 & 0.012&0.022 \\
$\Gamma$ & 25 & 25& 40\\
R [cm]&2.1e16 &8e16 & 2.1e16\\
$\theta_{obs}$ [degrees] & 2.3 &2.3 &1.4 \\ 
\hline
$L_e$ [erg s$^{-1}$]&4.6$\times10^{43}$ &2.3$\times10^{44}$& 1.4$\times10^{44}$\\
$L_B$ [erg s$^{-1}$]& 1.0$\times10^{43}$&2.2$\times10^{42}$ & 1.3$\times10^{42}$ \\
$L_B/L_e$ & 0.22& 9.4$\times10^{-3}$& 9.3$\times10^{-3}$\\
$t_{var}$ [hr] & 8.2& 31 &5.1 
\enddata
\end{deluxetable}

\end{document}